\documentclass[aps,pra,twocolumn,showpacs,superscriptaddress]{revtex4}
\usepackage{float,amssymb,amsfonts,amsmath,amsbsy,mathrsfs,graphicx,subfigure,bm,dcolumn,pifont,txfonts}

\newcommand{\ucite}[1]{\textsuperscript{\cite{#1}}}
\renewcommand{\eqref}[1]{Eq.\ (\ref{#1})}
\begin{document}
\title{New approach for solving master equation of open atomic system\footnote{Granted by Nation}}
\date{\today}
\author{Yi-Chong, Ren}
\email{rych@mail.ustc.edu.cn}
\affiliation{University of Science and Technology of China, Hefei, People's Republic of China}
\author{Hong-Yi, Fan}
\affiliation{University of Science and Technology of China, Hefei, People's Republic of China}
\affiliation{Shanghai Jiao Tong University, Shanghai, People's Republic of China}

\begin{abstract}
We describe a new approach called Ket-Bra Entangled State (KBES) Method which
enables one convert master equations into  Schr\"{o}dinger-like equation. In sharply contrast to the super-operator method, the KBES method is applicable for any master equation of
finite-level system in theory, and the calculation can be completed
by computer. With this method, we obtain the exact dynamic evolution of a radioactivity
damped $2$-level atom in time-dependent external field, and a $3$-level atom
coupled with bath; Moreover, the master equation of $N$-qubits Heisenberg chain each qubit coupled with a reservoir is also resolved in Sec.III; Besides, the paper briefly discuss the physical implications of the solution.
\end{abstract}

\maketitle

\section{Introduction}

The theory of open quantum systems which describes the dynamic evolution of
system coupled with environment is a fundamental approach to understand
dissipation and decoherence in quantum optics\ucite{r1}. One of the most
important instruments of open quantum systems is represented by master
equation\ucite{r2}, which plays a significant role in fundamental aspects, like
decoherence\ucite{r3,s1}, disentanglement\ucite{r4,s2}, quantum
dynamics\ucite{r5}, non-equilibrium thermodynamics\ucite{r6}, and applied ones
like atom emission\ucite{r7}, quantum transport\ucite{r8}, Bose-Einstein
condensates, Brownian motion\ucite{r6}, laser field\ucite{r11}.

Recently one topic has mostly drawn attention is the derivation, solution, and
systemic theory analysis of master equation which is a set complicated
differential equations of reduced density operator. Generally, the master
equations are solved by means of phase space representation of density
operator which can transform master equation into $c$-number equation such as
Fokker-Planck equation and Langevin equation\ucite{r12}. There is also some
other way like super-operator\ucite{r13,r14} method, which is
applicable only for that the Lindblad can be expressed in super-operator
generators of Su(1,1), Su(2)\ucite{r19,r20} and some other Lie algebra. We shall briefly review their method and compare it with our new method in Sec.II

In this contribution, a new procedure called Ket-Bra Entangled State(KBES)
method have been developed\ucite{r22}, which allow one to transform conventional master
equation into Schr\"{o}dinger-like equation, then the time evolution of density operator can be deduced; We can obtain the solution of Schr\"{o}dinger-like equation(see Sec.II). This method is applicable for any
master equation of finite-level system in theory. With this KBES method, we obtain the exact
solutions of three different physical models in Sec.III, which include a qubit
coupled with reservoir in time-dependent external field, a qutrit  with
Bose-bath, and a $N$-qubit Heisenberg chains coupled with reservoir.

\section{Super-Operator Method Versus KBES Method}

In this section, after briefly reviewing the general super-operator method, we
introduce Fan's method that constructs bosonic thermal entangled representation to mutual
transform general operator between real and fictitious mode\ucite{r18,r21}; As a generalization and
development of previous, we put forward KBES method which can convert
operator master equation into Schr\"{o}dinger-like equation.

To explain the super-operator method\ucite{r13,r14,r15}, consider an usual master equation%
\begin{equation}
\dot{\rho}=\mathscr{L}\rho, \label{d1}%
\end{equation}
whose Lindblad operator is
\begin{align}
\mathscr{L}\rho=  &  \frac{\gamma}{2}\left(  n+1\right)  \left(  2\sigma^{-}%
\rho\sigma^{+}-\sigma^{+}\sigma^{-}\rho-\rho\sigma^{+}\sigma^{-}\right)
\nonumber\\
&  +\newline\frac{\gamma}{2}n\left(  2\sigma^{+}\rho\sigma^{-}-\sigma
^{-}\sigma^{+}\rho-\rho\sigma^{-}\sigma^{+}\right)  . \label{d9}%
\end{align}
If $\mathscr{L}$ is time independent, the form solution to \eqref{d1} is%
\begin{equation}
\rho\left(  t\right)  =e^{\mathscr{L}t}\rho\left(  0\right)  . \label{d2}%
\end{equation}
$\rho\left(  t\right)  $ has the explicit expression if $\mathscr{L}$ consist of super-operator generators of Lie algebras, and there is a wide calss of master equation whose $\mathscr{L}$ can be expressed in terms of super-operator generators of Su(2) or Su(1,1) Lie algebra.

Previous literatures have defined the super-operators
\begin{equation}%
\begin{array}
[c]{l}%
L_{+}\rho=\sigma^{+}\rho\sigma^{-},L_{-}\rho=\sigma^{-}\rho\sigma^{+},\\
L_{z}\rho=\frac{1}{2}\left(  \sigma^{+}\sigma^{-}\rho-\rho\sigma^{-}\sigma
^{+}\right)  .
\end{array}
\label{d3}%
\end{equation}
where $\left[  L_{+},L_{-}\right]  =2L_{z},\left[  L_{z},L_{\pm}\right]  =\pm L_{\pm
}$ obeying Su(2) Lie algebra.
The explicit expression of $\mathscr{L}$ is
\begin{equation}
\mathscr{L}=-\frac{1}{2}\left(  2n+1\right)  \gamma+n\gamma L_{-}+\left(
n+1\right)  \gamma L_{+}-\gamma L_{z}. \label{d5}%
\end{equation}
Thus \eqref{d2} can be represented as%
\begin{align}
\rho\left(  t\right)   &  =e^{-\frac{1}{2}\left(  2n+1\right)  \gamma
t+n\gamma L_{-}+\left(  n+1\right)  \gamma tL_{+}-\gamma tL_{z}}\rho\left(
0\right)  ,\nonumber\\
&  =e^{x_{+}\left(  t\right)  L_{+}}e^{\ln x_{z}\left(  t\right)  L_{z}%
}e^{x_{-}\left(  t\right)  L_{-}}\rho\left(  0\right)  , \label{d6}%
\end{align}
where $x_{\pm}\left(  t\right)  ,x_{z}\left(  t\right)  $ is given in Ref.\ \cite{r19,r20}. Finally, the Kraus operator solution of \eqref{d6} is obtained by
transforming the super operator $L_{+},L_{-},L_{z}$ into gereral operator%
\begin{equation}
\rho\left(  t\right)  =\sum_{i}K_{i}\rho\left(  0\right)  K_{i}^{\dag},
\label{d7}%
\end{equation}
$K_{i}$ is the Kraus operator and satisfy $\sum_{i}K_{i}^{\dag}K_{i}=I$, where
$K_{i}$ is given by%
\begin{equation}%
\begin{array}
[c]{l}%
K_{1}=\frac{e^{-\left(  2n+1\right)  \gamma/4}}{\sqrt[4]{x_{z}}}x_{+}\left(
t\right)  \sigma^{+},\\
K_{2}=\frac{e^{-\left(  2n+1\right)  \gamma/4}}{\sqrt[4]{x_{z}}}x_{-}\left(
t\right)  \sigma^{-},\\
K_{3}=\frac{e^{-\left(  2n+1\right)  \gamma/4}}{\sqrt[4]{x_{z}}}x_{+}\left(
t\right)  x_{-}\left(  t\right)  \sigma^{+}\sigma^{-},\\
K_{4}=\frac{e^{-\left(  2n+1\right)  \gamma/4}}{\sqrt[4]{x_{z}}}\left[
I+\sigma^{+}\sigma^{-}\left(  \sqrt{x_{z}}-1\right)  \right]  .
\end{array}
\label{d8}%
\end{equation}

Ulteriorly, prof. Fan\ucite{r18,r21} constructs the thermal entangled state
$|I\rangle=e^{a^{\dag}\tilde{a}^{\dag}}|0,\tilde{0}\rangle$ by introducing an
extra fictitious mode, for the Bose operator
\begin{equation}
a|I\rangle=\tilde{a}^{\dag}|I\rangle,\text{ \ }a^{\dag}|I\rangle=\tilde
{a}|I\rangle, \label{d11}%
\end{equation}
thus, for the super operator $L_{\pm}$ that consist of creation and annihilation
operator
\begin{equation}%
\begin{array}
[c]{c}%
L_{-}\rho|I\rangle\equiv a\rho a^{\dag}|I\rangle=a\tilde{a}\rho|I\rangle
\Rightarrow L_{-}=a\tilde{a},\\
L_{+}\rho|I\rangle\equiv a^{\dag}\rho a|I\rangle=a^{\dag}\tilde{a}^{\dag}%
\rho|I\rangle\Rightarrow L_{+}=a^{\dag}\tilde{a}^{\dag},
\end{array}
\label{d12}%
\end{equation}
The thermal entangled state $|I\rangle$ can transform general operator into fictitious mode, with the characteristic it be used to slove master
equation of Bose system.

Next, we shall introduce our new method, consider a density operator $\rho$ in
Hilbert space $\mathscr{H}$
\begin{equation}
\rho={\sum\limits_{m,n}}\rho_{m,n}\vert m\rangle\langle n\vert, \label{d13}%
\end{equation}
where $\vert m\rangle$ constitutes any complete orthogonal basis in
$\mathscr{H}$. By introducing an extra fictitious mode, we construct the
Ket-Bran entangled states%

\begin{equation}
|\eta\rangle={\sum\limits_{m}}|m,\tilde{m}\rangle\label{d14}%
\end{equation}
fot the density operator $\rho$
\begin{equation}
|\rho\rangle=\rho|\eta\rangle=%
{\displaystyle\sum\limits_{m,n}}
\rho_{m,n}|m,\tilde{n}\rangle. \label{d15}%
\end{equation}
With the defining of $| \eta\rangle$, 
for any operator$A_{mn}\equiv\left\vert m\right\rangle \left\langle n\right\vert $ in $\mathscr{H}$
\begin{equation}
A_{mn}\left\vert \eta\right\rangle 
=\left\vert m,\tilde{n}\right\rangle
=\left\vert \tilde{n}\right\rangle \left\langle \tilde{m}\right\vert
\left\vert \eta\right\rangle =\tilde{A}_{mn}^{\dag}\left\vert \eta
\right\rangle . \label{d16}%
\end{equation}
Besides, \eqref{d16} is valid for any $A\equiv {\sum_{m,n}}\mathfrak{a}_{mn}A_{mn}$ where $\mathfrak{a}_{mn}$ is real, namely $A\left\vert \eta\right\rangle={A}^{\dag}\left\vert \eta\right\rangle$.
\eqref{d16} show that the general operator can
also be transformed between real mode and fictitious mode by KBES just as Fan's method. 

Obviously the fictitious mode indeed represent the bra vector of system, so we
called $|\eta\rangle$ Ket-Bra Entangled State (KBES). Density operator $\rho$ can be translated into pure state $|\rho\rangle$ and general operator can mutual transform between real and fictitious
mode by KBES $|\eta\rangle$, which enable one to convert master equation into Schr\"{o}dinger-like equation.

The case of \eqref{d1} has been solved by KBES method in our previous work
\ucite{r22}, here we consider the Lindblad equation, which is most general form of Markovian and time-homogeneous master equation describing non-unitary evolution of the density matrix $\rho$, that is
\begin{align}
\frac{d\rho}{dt}=  &  -\frac{i}{\hbar}\left[  H,\rho\right]  +\sum
_{n,m=1}^{N^{2}-1}h_{n,m}[L_{n}\rho L_{m}^{\dagger}\nonumber\\
&  -\frac{1}{2}\left(  \rho L_{m}^{\dagger}L_{n}+L_{m}^{\dagger}L_{n}%
\rho\right)  ], \label{d17}%
\end{align}
where $H$ is Hamiltonian part, $L_{m}$ is arbitrary linear operators in system's Hilbert space, and $h_{n,m}$ is constant. With the KBES $|\eta_{L}\rangle=\sum_{n=1}^{N}|n,\tilde{n}\rangle$, \eqref{d17} can be transformed into
Schr\"{o}dinger-like equation
\begin{align}
\frac{d}{dt}|\rho\rangle &  =-\frac{i}{\hbar}\left\{  \left(  H-\tilde
{H}\right)  +\mathscr{L}_{lin}\right\}  |\rho\rangle,\nonumber\\
&  \equiv\mathscr{F}_{L}|\rho\rangle, \label{d18}
\end{align}
where
\begin{equation}
\mathscr{L}_{lin}=\sum_{n,m=1}^{N^{2}-1}h_{n,m}\left[  L_{n}\tilde{L}%
_{m}-\frac{1}{2}\left(  \tilde{L}_{n}^{\dagger}\tilde{L}_{m}+L_{m}^{\dagger
}L_{n}\right)  \right]  . \label{d19}%
\end{equation}
\

Schr\"{o}dinger-like \eqref{d18} can be solved by two different approaches. The
first is the evolution operator method. assume $\mathscr{F}_{L}$ is
independent of time $t$, then%
\begin{equation}
|\rho\left(  t\right)  \rangle=e^{\mathscr{F}_{L}t}|\rho\left(  0\right)
\rangle. \label{d20}%
\end{equation}
where the explicit matrix represent of $\mathscr{F}_{L}$ is demanded, then as mentioned before, the calculation of $e^{\mathscr{F}_{L}t}$ need to diagonalize a $N^{2}$-order matrix $\mathscr{F}_{L}$, the calculation is complicate but can be done by computer. For few simple
situations, we can also decompose $e^{F_{L}t}$ into several exponential form of operators with Lie algebra (see \eqref{d6}) just like super operator method.

The second approach is the stationary state method. For some high-dimensional
systems the calculation of $e^{\mathscr{F}_{L}t}$ may be difficult even with
computer, thus we calculate the eigenstates and eigenvalues of
$\mathscr{F}_{L}$%
\begin{equation}
\mathscr{F}_{L}\vert\varphi_{i}\rangle=\lambda_{i}\vert\varphi_{i}\rangle.
\label{d21}%
\end{equation}
Then the solution can be represented as follow:
\begin{equation}
\vert\rho\left(  t\right)  \rangle=%
{\displaystyle\sum\limits_{i}}
C_{i}e^{\lambda_{i}t}\vert\varphi_{i}\rangle, \label{d22}
\end{equation}
where $C_{i}$ is constant which can be determined by initially $\rho\left(0\right)  $ and characters of density operator. 
{\bf Obviously, all eigenvalues $\lambda_{i}\leq0$ and the eigenstate $\vert\varphi\rangle$ whose eigenvalue $\lambda=0$ corresponding the final state $\rho\left(  \infty\right)  $, For $t\rightarrow\infty$, only the eigenvector term whose eigenvalue is zero left, other term's coefficient $C_{i}e^{\lambda_{i}}$ shall disappear along with the growth of $t$, this characteristic can obtain the final density operator directly.} Furthermore, most methods of Schr\"{o}dinger equation previously can also be used to solve master equation with KBES.

Super-operator method is concise and the solution is applicable for any initial state, however restricted by Lie algebra, it show a narrow applicable range, and even slightly changes of master equation may lead to unsolvable effect for super operator method.
Compared with the both method, our KBES method have three merits: {\bf1. The procedure is applicable for any master equation of
finite-level systems in theoretical; 2. The process of resolution is most
concise and can be fully completed by computer; 3. The method can convert
master equation into Schr\"{o}dinger-like equation, which means most methods
of Schr\"{o}dinger equation can be used to solve master equation. These ascendances shall be proved in the process of solving follow models.}
\section{Several Physical Models}

To further concretely explain our KBES method, we shall introduce three different physical models in this section, moreover each model can't be solved by general super-operator method. Especially the third model, to best of our knowledge is still unsolved for large $N$, whereas all these can be solved by KBES method.


\subsection{Damped 2-level Atom Model}

Consider an atom in cavity full of external field that is radiatively damped
by its interaction with the various modes of bath in thermal equilibrium at
temperature $T$ just as Fig.1. \begin{figure}[h]
\centering
\includegraphics[width=0.25\textwidth]{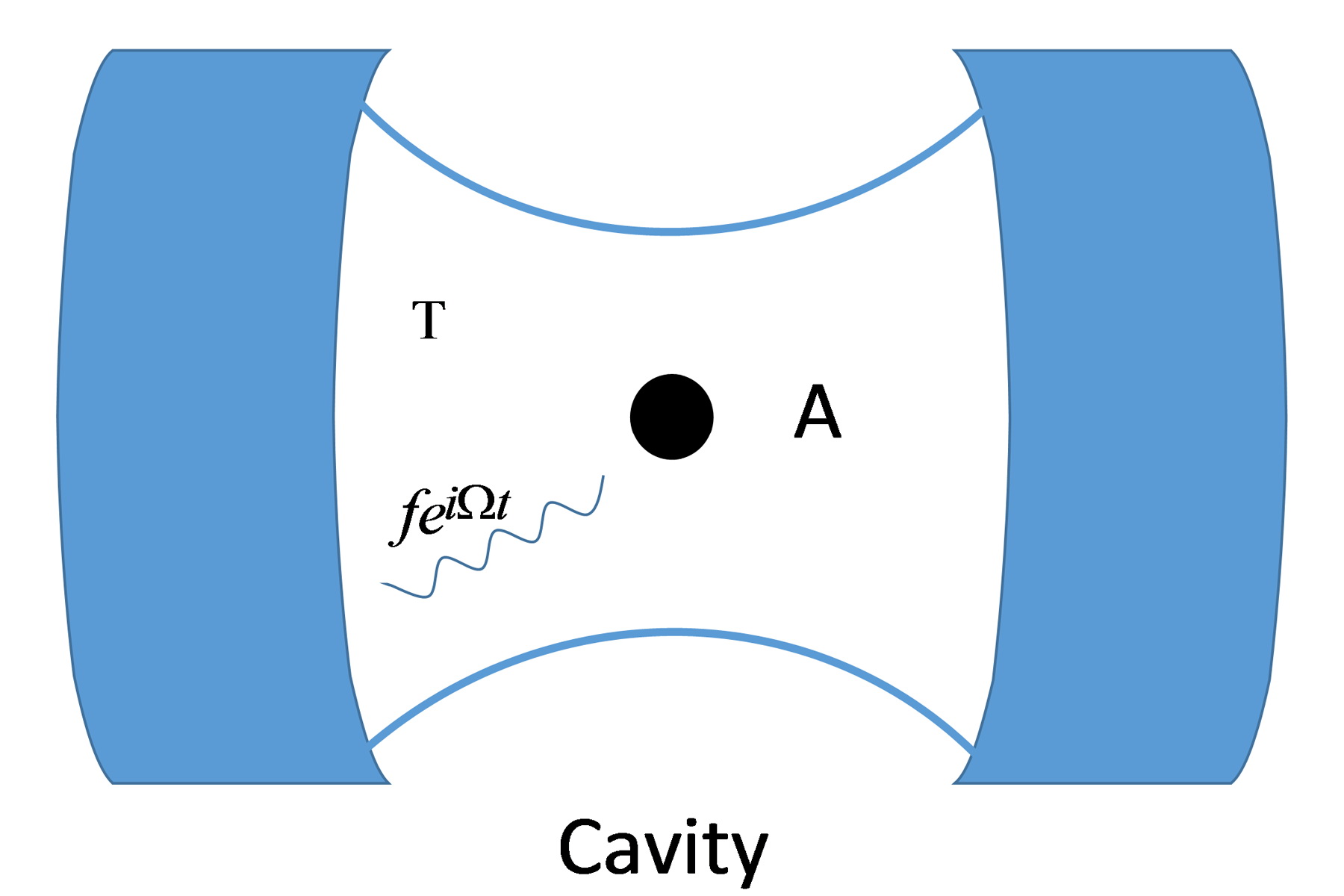}\caption{Qubit with Cavity
Damping in External Field}%
\end{figure}The Hamiltonian is given in the rotating-wave and dipole
approximations as follow:%
\begin{align}
H  &  =\omega_{0}\sigma_{z}+%
{\displaystyle\sum\limits_{k}}
\omega_{k}a_{k}^{\dagger}a_{k}+f\left(  \sigma^{+}e^{i\Omega t}+\sigma
^{-}e^{-i\Omega t}\right) \nonumber\\
&  +\sigma^{-}%
{\displaystyle\sum\limits_{k}}
g_{k}^{\ast}a_{k}^{\dagger}+\sigma^{+}%
{\displaystyle\sum\limits_{k}}
g_{k}a_{k}, \label{d23}%
\end{align}
where $f$ and $\Omega$ represent the amplitude and frequency of
external field respectively. In the interaction picture%
\begin{align}
H_{I}  &  =e^{iH_{0}t}\left(  H-H_{0}\right)  e^{-iH_{0}t}\nonumber\\
&  =H^{\prime}+H_{I}^{\prime}, \label{d24}%
\end{align}
where $\Lambda=\omega_{0}-\Omega$ and
\begin{align}
H_{0}  &  =\Omega\sigma_{z}+%
{\displaystyle\sum\limits_{k}}
\omega_{k}a_{k}^{\dagger}a_{k}\nonumber\\
H^{\prime}  &  \equiv\Lambda\sigma_{z}+f\left(  \sigma^{+}+\sigma_{-}\right)
\nonumber\\
H_{I}^{\prime}  &  =\sigma^{-}e^{i\Omega t}%
{\displaystyle\sum\limits_{k}}
g_{k}^{\ast}a_{k}^{\dagger}e^{i\omega_{k}t}+\sigma^{+}e^{-i\Omega t}%
{\displaystyle\sum\limits_{k}}
g_{k}a_{k}e^{-i\omega_{k}t}. \label{d25}%
\end{align}
Through the derivation of master equation in Ref. \ucite{r2}, we can obtain the
corresponding master equation of this model as follow

\begin{align}
\frac{d\rho}{dt}  &  =-i\left[  H^{\prime},\rho\right]  +\frac{\gamma}%
{2}n\left(  2\sigma^{+}\rho\sigma^{-}-\sigma^{-}\sigma^{+}\rho-\rho\sigma
^{-}\sigma^{+}\right) \nonumber\\
&  +\frac{\gamma}{2}\left(  n+1\right)  \left(  2\sigma^{-}\rho\sigma
^{+}-\sigma^{+}\sigma^{-}\rho-\rho\sigma^{+}\sigma^{-}\right)  , \label{d26}%
\end{align}
where $H^{\prime}$ is the entangled Hamiltonian of qubit and external field,
$\gamma$ is the spontaneous emission rate and $n=n\left(  \Omega,T\right)  $
is the photon number of frequency $\Omega$ in temperature $T$.

With the corresponding KBES $|\eta_{Q}\rangle=|0,\tilde{0}\rangle+|1,\tilde
{1}\rangle$, we can deduce the Schr\"{o}dinger-like equation of \eqref{d26}:%
\begin{equation}
\frac{d}{dt}|\rho\rangle=\mathscr{F}|\rho\rangle, \label{d27}%
\end{equation}
where $\displaystyle\alpha\equiv\gamma\left(  n+1\right)  /2,\beta
\equiv\gamma n/2$ and
\begin{align}
\mathscr{F}=  &  i\left(  \tilde{H}^{\prime}-H^{\prime}\right)  +\alpha\left(
2\sigma^{-}\tilde{\sigma}^{-}-\sigma^{+}\sigma^{-}-\tilde{\sigma}^{+}%
\tilde{\sigma}^{-}\right) \nonumber\\
&  +\beta\left(  2\sigma^{+}\tilde{\sigma}^{+}-\sigma^{-}\sigma^{+}%
-\tilde{\sigma}^{-}\tilde{\sigma}^{+}\right)  . \label{d28}%
\end{align}
The formal solution is $\displaystyle|\rho\left(  t\right)  \rangle
=e^{\mathscr{F}t}|\rho\left(  0\right)  \rangle$, and the explicit formulation
of $\mathscr{F}$ in Kronecker product space can be given: {\small
\begin{equation}
\mathscr{F}=\left(
\begin{array}
[c]{cccc}%
-2\alpha & if & -if & 2\beta\\
if & -\alpha-\beta-2i\Lambda & 0 & -if\\
-if & 0 & -\alpha-\beta+2i\Lambda & if\\
2\alpha & -if & if & -2\beta
\end{array}
\right)  , \label{d29}%
\end{equation}
}where $\displaystyle\sigma^{_{\pm}}=\sigma^{_{\pm}}\otimes\tilde{I}%
,\tilde{\sigma}^{_{\pm}}=I\otimes\tilde{\sigma}^{_{\pm}}$. In fact,
$\displaystyle e^{\mathscr{F}t}$ can be obtained through the diagonalization of \eqref{d29}, however the explicit expression of $\rho\left(  t\right)  $ is too long and complicated to be presented in here. As we have mentioned in Sec.II, {\bf the final density $\rho\left(  \infty\right)  $ is represented by the eigenvector of $\mathscr{F}$ whose corresponding eigenvalue equal to zero.} The eigenvector whose eigenvalue equals to zero is
\begin{equation}
\left\vert \varphi\right\rangle =\left(
\begin{array}
[c]{c}
\frac{\left(  f^{2}+\beta(\alpha+\beta)\right)  +4\Lambda^{2}\beta/\left(
	\alpha+\beta\right)  }{2\left(  f^{2}+2\Lambda^{2}\right)  +\left(
	\alpha+\beta\right)  ^{2}}\\
\frac{-if(\alpha-\beta)(\alpha+\beta-2i\Lambda)}{2\left(  f^{2}+2\Lambda
	^{2}\right)  +\left(  \alpha+\beta\right)  ^{2}}\\
\frac{if(\alpha-\beta)(\alpha+\beta-2i\Lambda)}{2\left(  f^{2}+2\Lambda
	^{2}\right)  +\left(  \alpha+\beta\right)  ^{2}}\\
\frac{\left(  f^{2}+\alpha(\alpha+\beta)\right)  +4\Lambda^{2}\alpha/\left(
	\alpha+\beta\right)  }{2\left(  f^{2}+2\Lambda^{2}\right)  +\left(
	\alpha+\beta\right)  ^{2}}%
\end{array}
\right).
\end{equation}
Thus $\rho\left(  \infty\right)$ is given by
\begin{equation}
\rho\left(  \infty\right)  =\frac{1}{2n+1}\left(
\begin{array}
[c]{cc}%
n+\frac{f_{\gamma}^{2}}{M\left(  n,f_{\gamma},\Lambda_{\gamma}\right)  } &
\frac{-if_{\gamma}(n+1/2-2i\Lambda_{\gamma})}{M\left(  n,f_{\gamma}%
,\Lambda_{\gamma}\right)  }\\
\frac{if_{\gamma}(n+1/2+2i\Lambda_{\gamma})}{M\left(  n,f_{\gamma}%
,\Lambda_{\gamma}\right)  } & \frac{\left(  n+1\right)  M\left(  n,f_{\gamma
},\Lambda_{\gamma}\right)  -f_{\gamma}^{2}}{M\left(  n,f_{\gamma}%
,\Lambda_{\gamma}\right)  }%
\end{array}
\right)  , \label{d30}%
\end{equation}
where $\displaystyle f_{\gamma}\equiv 2f/\gamma,\Lambda_{\gamma}\equiv
2\Lambda/\gamma$ and
\begin{equation}
M\left(  n,f_{\gamma},\Lambda_{\gamma}\right)  =(n+1/2)^{2}+2\left(
f_{\gamma}^{2}+2\Lambda_{\gamma}^{2}\right)  . \label{d31}%
\end{equation}

Plot the figure of $\rho_{11}\left(  t\right)  $ verus $\gamma t$ and
$\displaystyle \Lambda_{\gamma}$ on different conditions with initial state
$\rho\left(  0\right)  =\vert\varphi\rangle\langle\varphi\vert, \vert
\varphi\rangle=\frac{1}{\sqrt{2}}\left(  \vert0\rangle+\vert1\rangle\right)  $
as follow:

\begin{figure}[h]
\setcounter{subfigure}{0} \centering
\subfigure[$f_{\gamma}=0,n=0$]{
\includegraphics[width=0.14\textwidth]{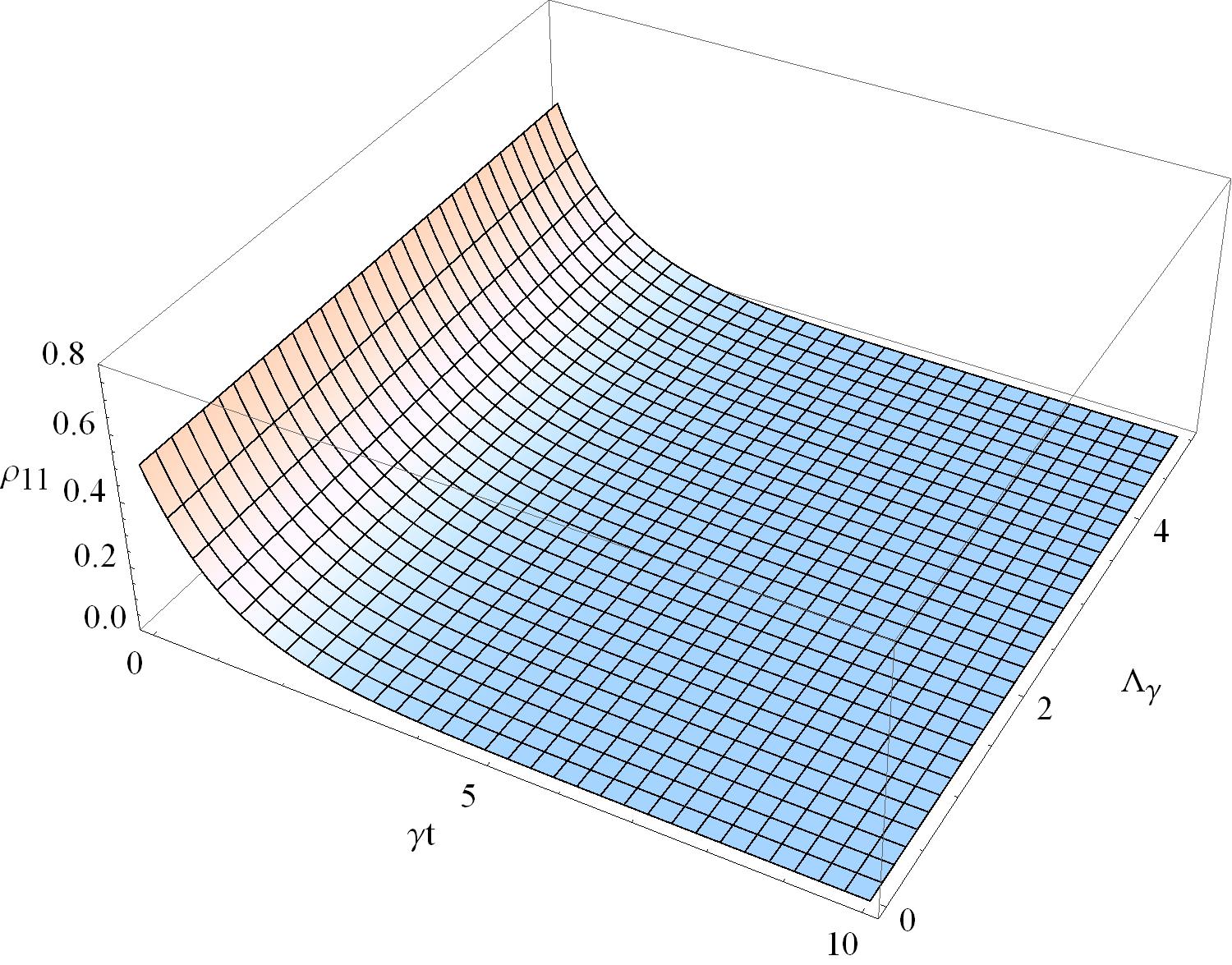}} \hspace{0.05in}
\subfigure[$f_{\gamma}=1,n=0$]{
\includegraphics[width=0.14\textwidth]{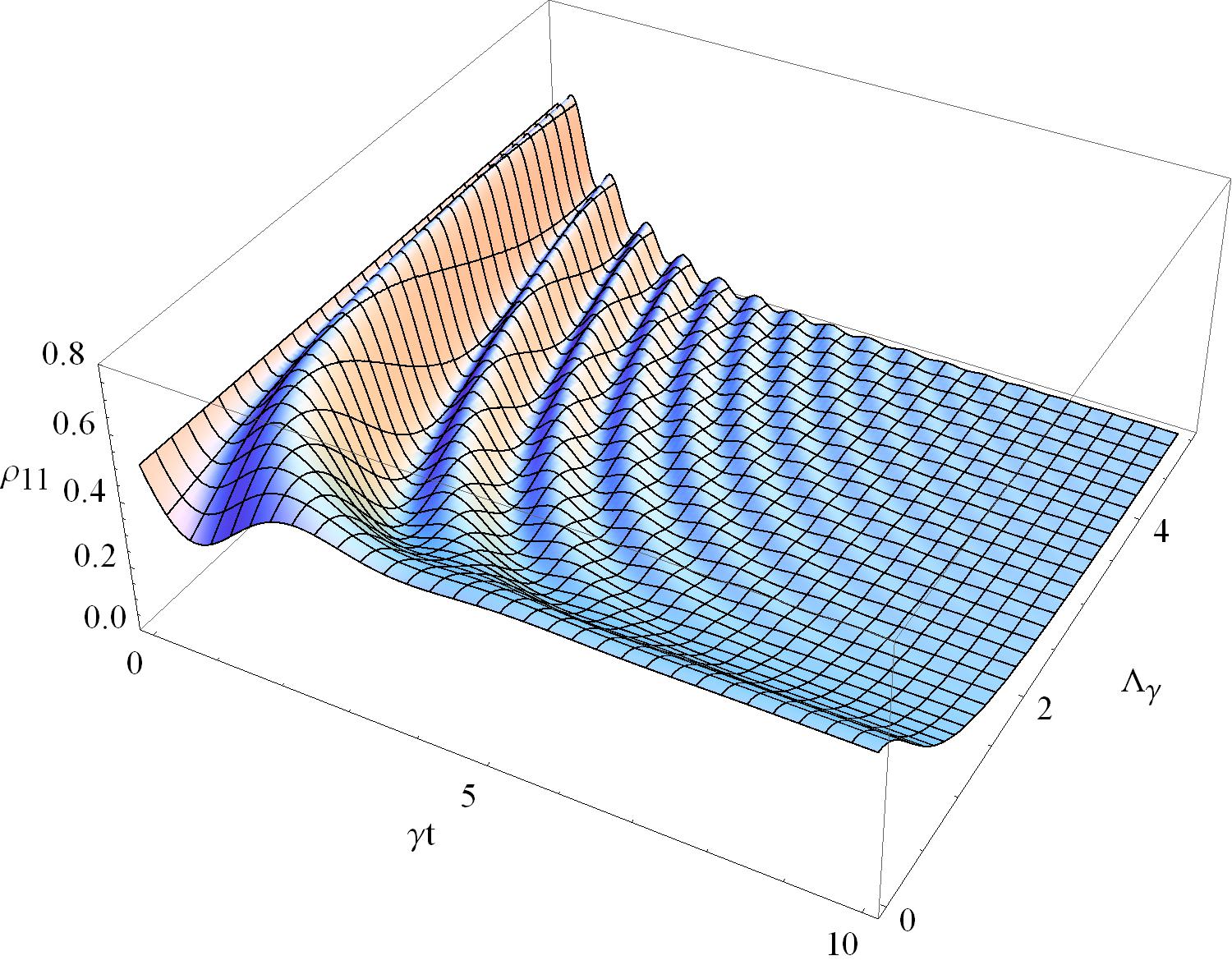}} \hspace{0.05in}
\subfigure[$f_{\gamma}=2,n=0$]{
\includegraphics[width=0.14\textwidth]{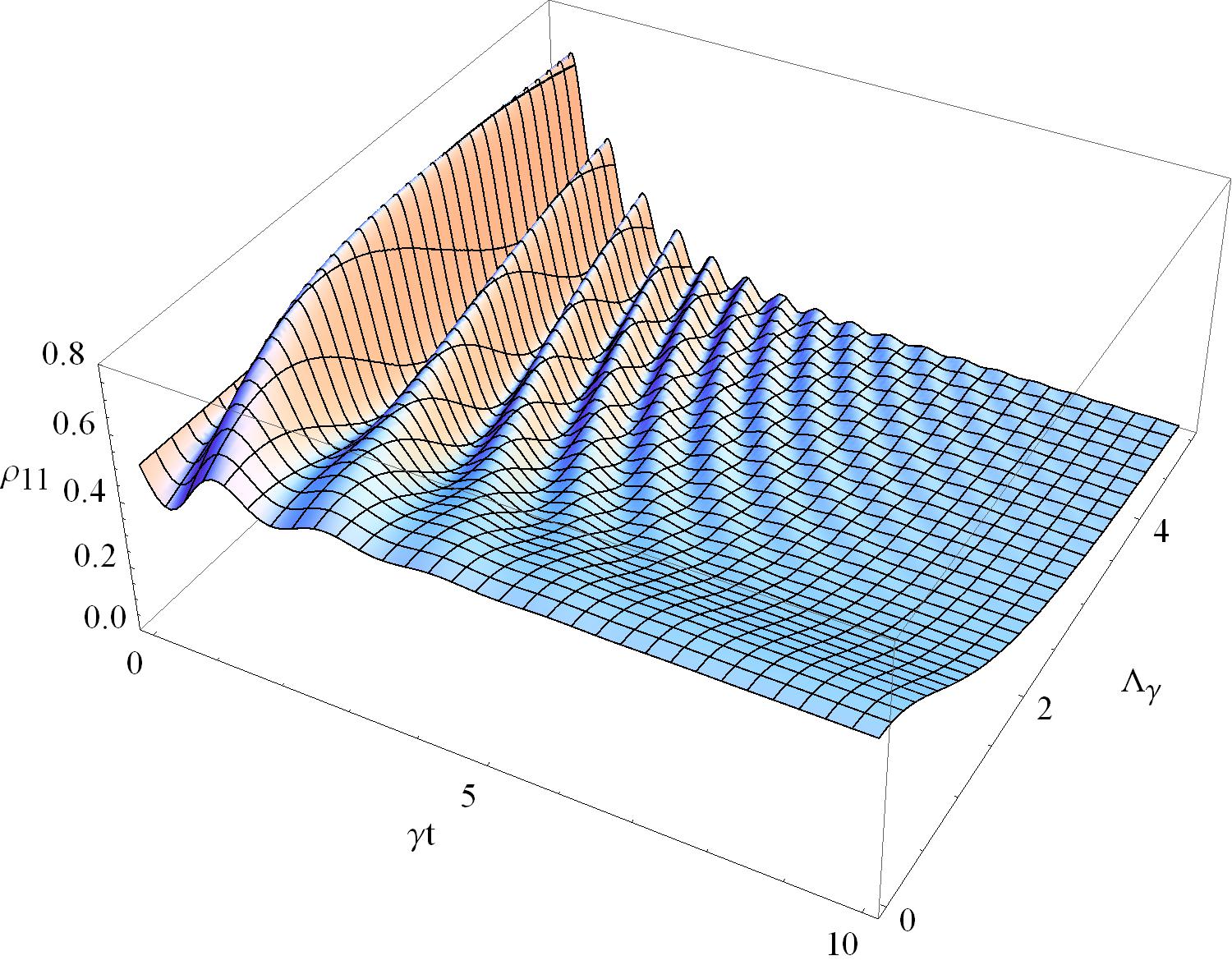}}\caption{$\rho_{11}%
\text{\ versus\ }\gamma t\text{\ and\ }\Lambda_{\gamma}.$}%
\end{figure}

Obviously, $\rho_{11}\left(  \infty\right)  $ that is severely influenced by
amplitude of external field $f$ and frequency $\Lambda$; For a fixed $\Lambda
$, $\rho_{11}\left(  \infty\right)  $ show a positive correlation with $f$,
namely $\rho_{11}\left(  \infty\right)  $ increases along with the
augmentation of $f$; However for a fixed $f>0$, there is a negative
correlation between $\rho_{11}\left(  \infty\right)  $ and $\vert\Lambda\vert$
which lead to a local maximum of $\rho_{11}\left(  \infty\right)  $ at
$\Lambda=0$; While for $f=0$, the change of $\Lambda$ make no difference to
$\rho_{11}\left(  t\right)  $. For any $f\neq0$, we find larger $\vert
\Lambda\vert$ shall lead to sharp and frequent fluctuations on the outiline of
$\rho_{11}\left(  t\right)  $.

While the figure of non-diagonal element $|\rho_{10}\left(  t\right)  |$ is:
\begin{figure}[h]
\setcounter{subfigure}{0} \centering
\subfigure[$f_{\gamma}=0,n=0$]{
\includegraphics[width=0.14\textwidth]{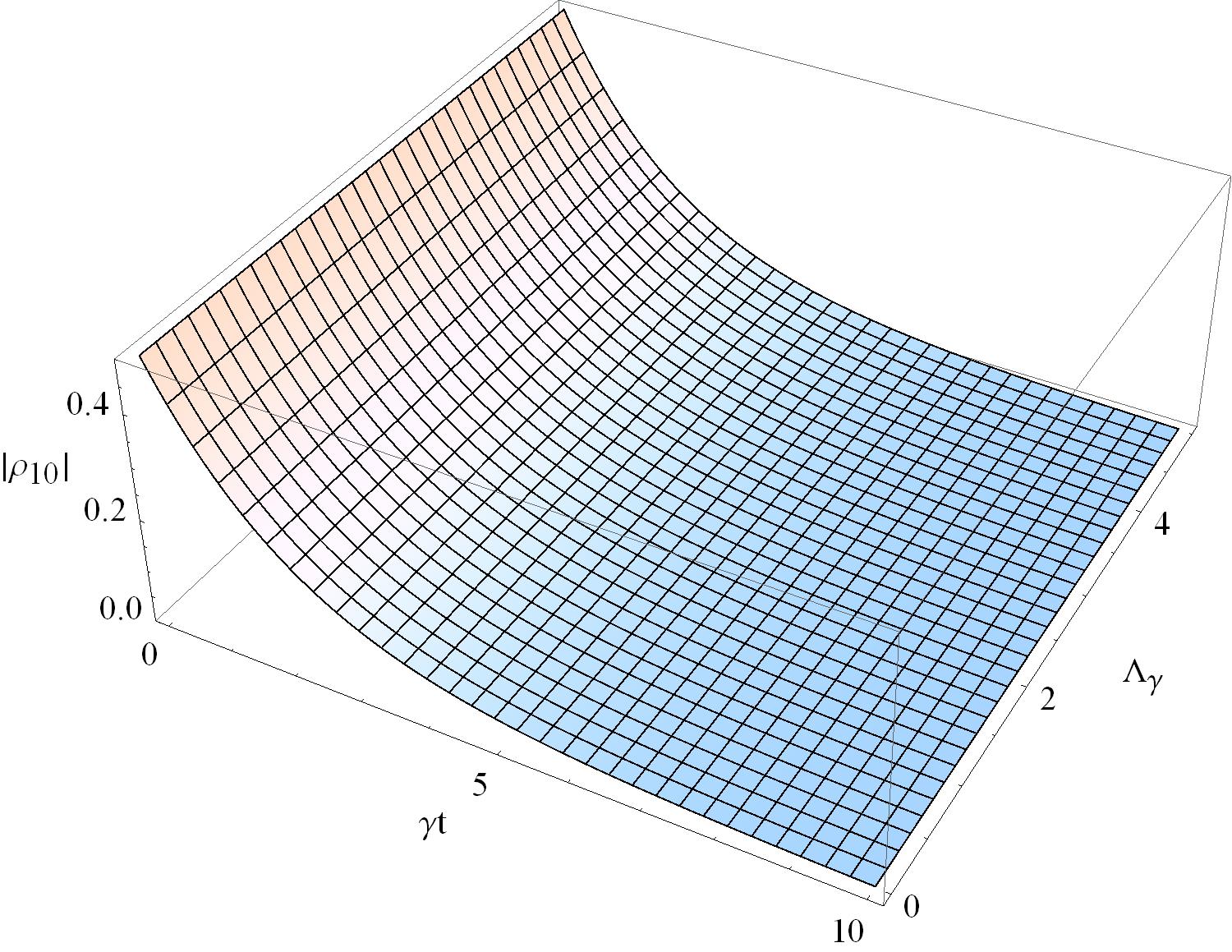}} \hspace{0.05in}
\subfigure[$f_{\gamma}=1,n=0$]{
\includegraphics[width=0.14\textwidth]{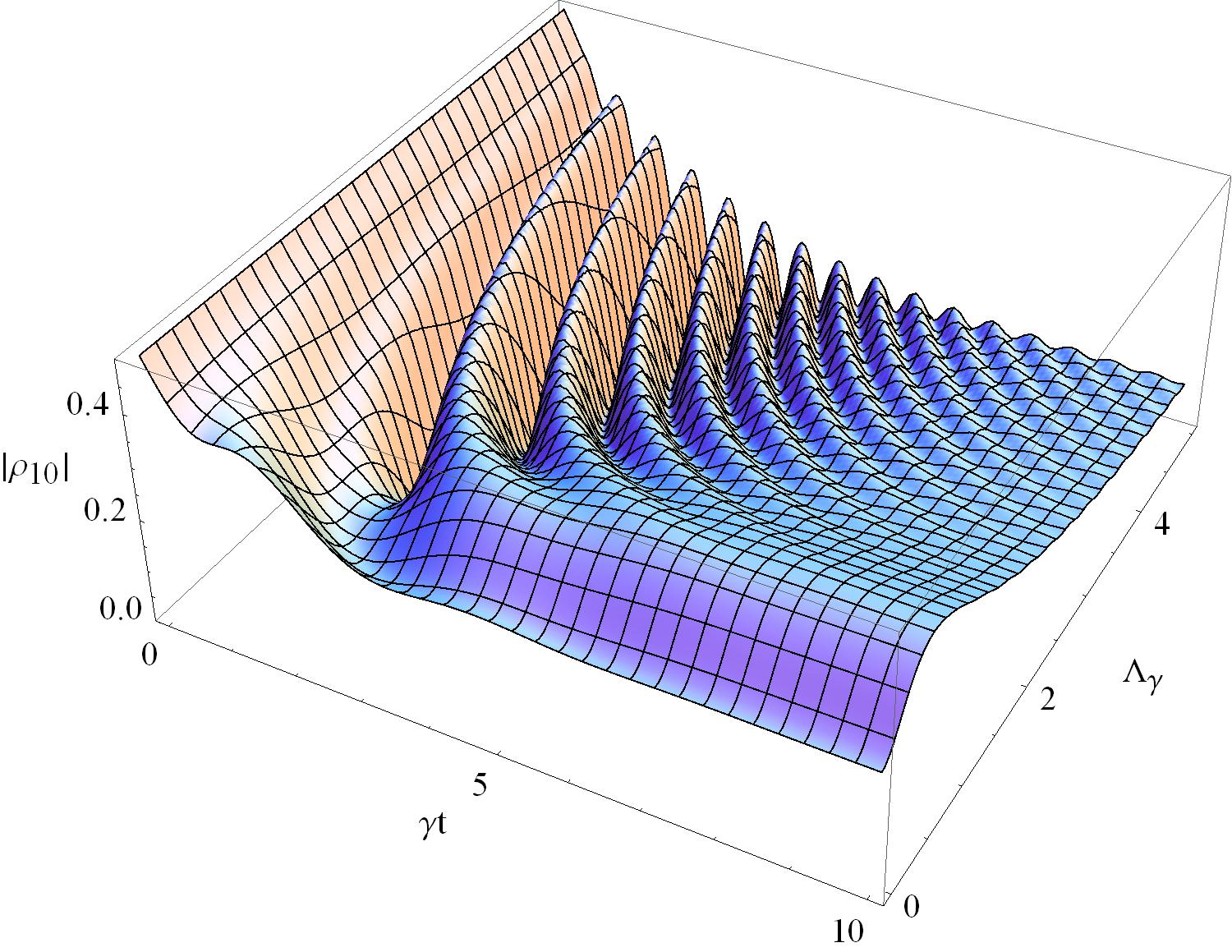}} \hspace{0.05in}
\subfigure[$f_{\gamma}=2,n=0$]{
\includegraphics[width=0.14\textwidth]{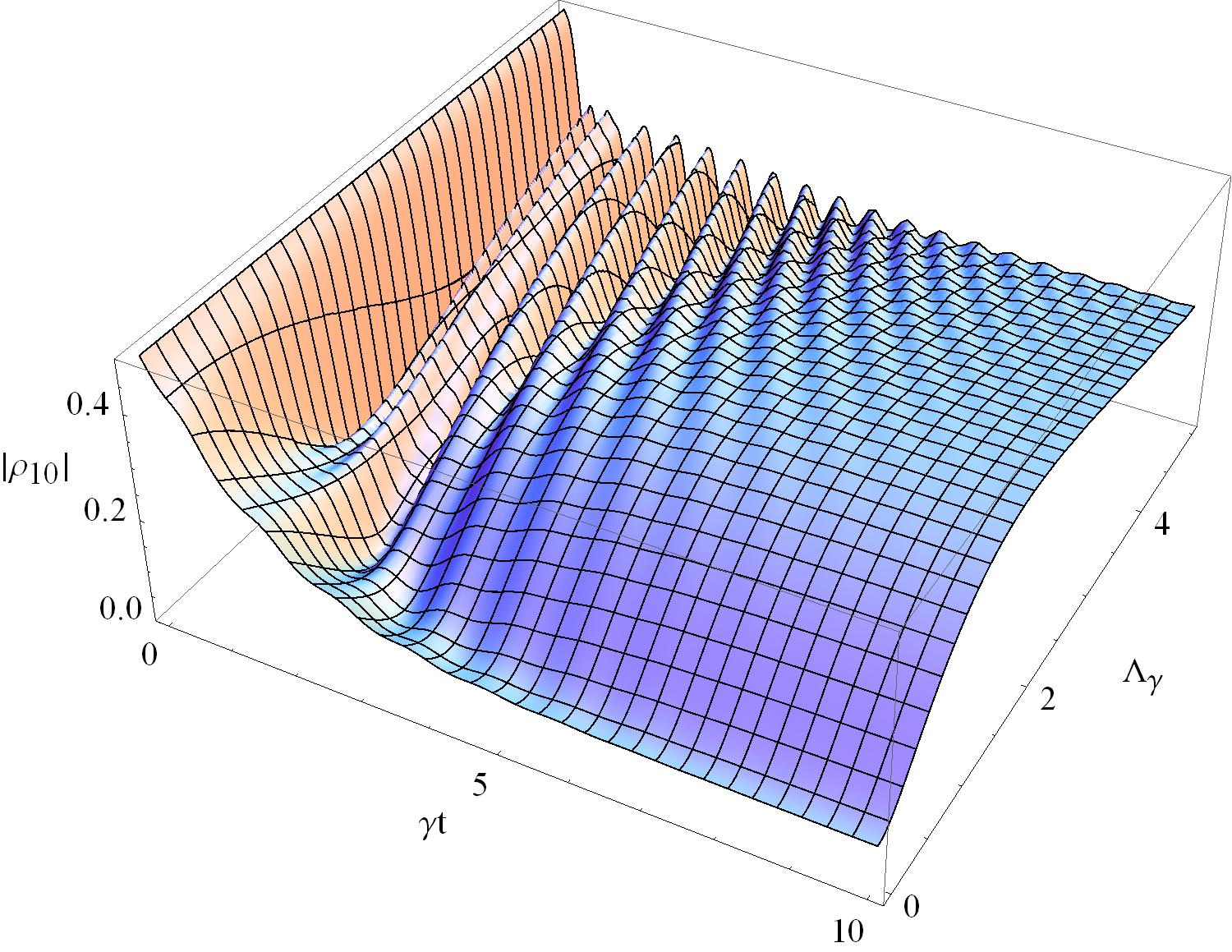}}\caption{$\left\vert \rho
_{10}\right\vert \text{\ versus\ }\gamma t\text{\ and\ }\Lambda_{\gamma}.$}%
\end{figure}\eqref{d30} show that $\displaystyle|\rho_{10}\left(  \infty\right)
|^{2}$ take the maximum
\begin{equation}
|\rho_{10}\left(  \infty\right)  |_{Max}^{2}=\frac{f_{\gamma}^{2}}{4\left(
n+1/2\right)  ^{2}-1+8f_{\gamma}^{2}}, \label{d32}%
\end{equation}
when $\displaystyle8\Lambda_{\gamma}^{2}+1=2\left(  n+1/2\right)
^{2}+4f_{\gamma}^{2}$. For zero temperature $\displaystyle n=0$, $|\rho
_{10}\left(  \infty\right)  |$ has a maximum $1/2\sqrt{2}$ at $16\Lambda
^{2}+\gamma^{2}=8f^{2}$, that has been presented in Fig.3. Despite these, for
$f=0$, change of $\Lambda$ make no difference and larger $|\Lambda|$ lead to
more frequent fluctuations on the outiline.

All those can be easily interpreted , $f$ represents the amplitude of external
field, so larger $f$ means strong external field which lead to the transition
of qubit from ground to excited state, i.e. the increasing of $\rho
_{11}\left(  \infty\right)  $ and decreasing of $\rho_{00}\left(
\infty\right)  $. When $\Lambda=0$ means the qubit and external field have a
same frequency $\omega_{0}=\Omega$, there is resonance between qubit and
external field, so $\rho_{11}\left(  \infty\right)  $ have a local maximum
value\ in here; The lack of space forbids a further detail discussion of the
off-diagonal elements $\rho_{10}\left(  t\right)  $ in here.

\subsection{$3$-level $V$-type Atom Model}

Now we extend the research of qubit to the case of $V$-type three-level
qutrit. Interesting example come from the three-level $V$-type atomic system
where spontaneous emission may take place from two excited levels to the
ground state but direct transition between excited levels is forbidden.
However the indirect coupling between excited states can appear due to
interaction with the ground state (quantum interference).

\begin{figure}[h]
\centering
\includegraphics[width=0.25\textwidth]{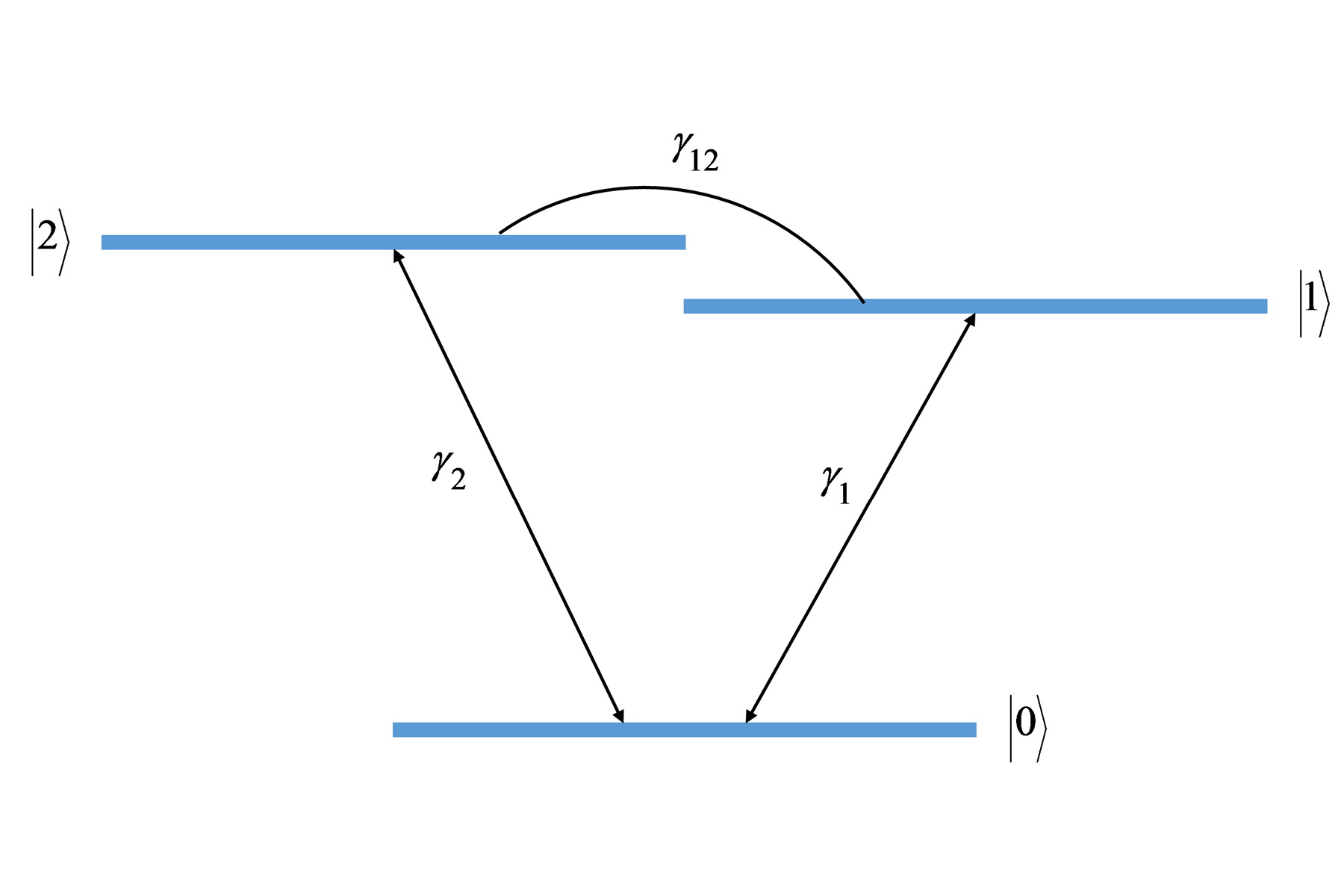}\caption{$V$-type Damped
Qutrit}%
\end{figure}

Consider a three-level V-type atom as in Fig.4, in which the atom has two
nondegenerate excited states $|1\rangle,|2\rangle$ with transition frequencies
to ground state $|0\rangle$ given by $\omega_{1},\omega_{2}$ and spontaneous
emission rate $\gamma_{1},\gamma_{2}$ respectively. The indirecting coupling
between $|1\rangle$ and $|2\rangle$ can represent as cross damping constant%
\begin{equation}
\gamma_{12}=\beta_{I}\sqrt{\gamma_{1}\gamma_{2}}, \label{d33}%
\end{equation}
where $\beta_{I}$ change from $0$ to $1$ represents the mutual orientation of
transition dipole moments.

The dynamics of such system is given by the master equation\ucite{s3,s4}%
\begin{align}
\frac{d\rho}{dt}=  &  \frac{\gamma_{1}}{2}\left(  2\sigma_{01}\rho\sigma
_{10}-\left\{  \sigma_{11},\rho\right\}  \right)  +\frac{\gamma_{2}}{2}\left(
2\sigma_{02}\rho\sigma_{20}-\left\{  \sigma_{22},\rho\right\}  \right)
\nonumber\\
&  +\frac{\gamma_{12}}{2}\left(  2\sigma_{01}\rho\sigma_{20}+2\sigma_{02}%
\rho\sigma_{10}-\left\{  \sigma_{12}+\sigma_{21},\rho\right\}  \right)  ,
\label{d34}%
\end{align}
where $\displaystyle\left\{  A,B\right\}  =AB+BA$ is the anticommutation and
$\sigma_{mn}$ $=|m\rangle\langle n|$ means the transition from $|n\rangle$ to
$|m\rangle$. Ref.\ \cite{s3,s4} solve analogous equation with the
conditions $\gamma_{12}=0$ and $\gamma_{1}=\gamma_{2}$. We shall solve this
master equation without any assumption and approximation in present work.
Construct the corresponding KBES%
\begin{equation}
|\eta_{T}\rangle=|0\tilde{0}\rangle+|1\tilde{1}\rangle+|2\tilde{2}\rangle,
\label{d35}%
\end{equation}
\eqref{d34} can be rewrite as Schr\"{o}dinger-like equation%
\begin{equation}
\frac{d}{dt}|\rho\rangle=\mathscr{F}_{T}|\rho\rangle, \label{d36}%
\end{equation}
where {\small
\begin{align}
\mathscr{F}_{T}=  &  \frac{\gamma_{1}}{2}\left(  2\sigma_{01}\tilde{\sigma}_{01}%
-\sigma_{11}-\tilde{\sigma}_{11}\right)  +\frac{\gamma_{2}}{2}\left(
2\sigma_{02}\tilde{\sigma}_{02}-\sigma_{22}-\tilde{\sigma}_{22}\right)
\nonumber\\
&  +\frac{\gamma_{12}}{2}\left(  2\sigma_{01}\tilde{\sigma}_{02}+2\sigma
_{02}\tilde{\sigma}_{01}-\sigma_{12}-\tilde{\sigma}_{21}-\sigma_{21}%
-\tilde{\sigma}_{12}\right)  . \label{d37}%
\end{align}
} Then the formal solution is:%
\begin{equation}
|\rho\left(  t\right)  \rangle=e^{\mathscr{F}_{T}}|\rho\left(  0\right)  \rangle.
\label{d38}%
\end{equation}

The explicit solution of \eqref{d38} can be given by diagonalizing $\mathscr{F}_{T}$ to
get the explicit expression of $e^{\mathscr{F}_{T}t}$, which can be finished by computer
with Mathematica. However the explicit expression is too complicated to be
given here, so we give the solution for $\gamma_{1}=\gamma_{2}=\gamma
,\gamma_{12}=\beta_{I}\gamma$, the five independent matrix elements is:
{\small
\begin{align}
\rho_{10}\left(  t\right)   &  =\frac{1}{2}e^{-\frac{1}{2}\left(  1+\beta
_{I}\right)  \gamma t}\left[  \rho_{10}+\rho_{20}+e^{\beta_{I}\gamma t}\left(
\rho_{10}-\rho_{20}\right)  \right]  ,\nonumber\\
\rho_{20}\left(  t\right)   &  =\frac{1}{2}e^{-\frac{1}{2}\left(  1+\beta
_{I}\right)  \gamma t}\left[  \rho_{10}+\rho_{20}+e^{\beta_{I}\gamma t}\left(
\rho_{20}-\rho_{10}\right)  \right]  ,\nonumber\\
\rho_{21}\left(  t\right)   &  =\frac{1}{2}e^{-\gamma t}\left[
\begin{array}
[c]{c}%
\rho_{21}-\rho_{12}+\left(  \rho_{12}+\rho_{21}\right)  e^{\beta_{I}\gamma
t}\\
-\left(  \rho_{11}+\rho_{22}\right)  \sinh\left(  \beta_{I}\gamma t\right)
\end{array}
\right]  ,\nonumber\\
\rho_{11}\left(  t\right)   &  =\frac{1}{2}e^{-\gamma t}\left[
\begin{array}
[c]{c}%
\rho_{11}-\rho_{22}+\left(  \rho_{11}+\rho_{22}\right)  \cosh\left(  \beta
_{I}\gamma t\right) \\
-\left(  \rho_{12}+\rho_{21}\right)  \sinh\left(  \beta_{I}\gamma t\right)
\end{array}
\right]  ,\nonumber\\
\rho_{22}\left(  t\right)   &  =\frac{1}{2}e^{-\gamma t}\left[
\begin{array}
[c]{c}%
\rho_{22}-\rho_{11}+\left(  \rho_{11}+\rho_{22}\right)  \cosh\left(  \beta
_{I}\gamma t\right) \\
-\left(  \rho_{12}+\rho_{21}\right)  \sinh\left(  \beta_{I}\gamma t\right)
\end{array}
\right]  . \label{d39}%
\end{align}
}Remark $\rho_{mn}=\rho_{mn}\left(  0\right)  $ in the equation for simplify,
other elements can be easily obtained with $\rho_{ij}\left(  t\right)
=\rho_{ji}^{\ast}\left(  t\right)  $ and $\rho_{00}\left(  t\right)
=1-\rho_{22}\left(  t\right)  -\rho_{11}\left(  t\right)  .$

\subsection{$N$-qubit $XXZ$ Heisenberg model}

Here we consider a $N$-qubit anisotropic $XYZ$ Heisenberg chain, the
Hamiltonian is given by%
\begin{equation}
H_{S}={\sum_{i=1}^{N}}\left[  J_{x}\sigma_{i}^{x}\sigma_{i+1}^{x}+J_{y}%
\sigma_{i}^{y}\sigma_{i+1}^{y}+J_{z}\sigma_{i}^{z}\sigma_{i+1}^{z}\right]  ,
\label{g1}%
\end{equation}
where $\sigma_{i}^{\varepsilon}\left(  \varepsilon=x,y,z\right)  $ are the
Pauli matrices of the $i$-th qubit. $J_{\varepsilon}\left(  \varepsilon
=x,y,z\right)  $ are the strengths of the spin interaction. For the
interaction, when $J_{x}=J_{y}\neq J_{z}$, the model can be called $XXZ$
chain. \eqref{g1} is rewriten as%
\begin{equation}
H_{S}={\sum\limits_{i}}\left[  J\left(  \sigma_{i}^{+}\sigma_{i+1}^{-}%
+\sigma_{i}^{-}\sigma_{i+1}^{+}\right)  +J_{z}\sigma_{i}^{z}\sigma_{i+1}%
^{z}\right]  , \label{g2}%
\end{equation}
in which $J=J_{x}+J_{y}$. The eigenvector of $H_{S}$ can be exactly solved by
Jordan-Wigner transformation\ucite{r23}. Now, we only consider the
"one-particle" eigenvector (i.e. only one qubit in spin up)%
\begin{equation}
\vert k\rangle=\frac{1}{\sqrt{N}}{\sum\limits_{n=1}^{N}}\exp\left(
\frac{i2\pi nk}{N}\right)  \sigma_{n}^{+}\vert0^{\otimes N}\rangle, \label{g3}%
\end{equation}
and the inverse transformation is%
\begin{equation}
\sigma_{n}^{+}\vert0^{\otimes N}\rangle=\frac{1}{\sqrt{N}}{\sum\limits_{k=1}%
^{N}}\exp\left(  \frac{-i2\pi nk}{N}\right)  \vert k\rangle. \label{g4}%
\end{equation}
For the vector $\vert k\rangle$ and $\vert0^{\otimes N}\rangle$, the
eigenequation is%
\begin{equation}
\left\{
\begin{array}
[c]{l}%
H_{S}\vert k\rangle=E_{k}\vert k\rangle,\\
H_{S}\vert0^{\otimes N}\rangle=E_{0}\vert0^{\otimes N}\rangle,
\end{array}
\right.  \label{g5}%
\end{equation}
the eigenvalues are%
\begin{equation}
\left\{
\begin{array}
[c]{l}%
E_{0}=NJ_{z},\\
E_{k}=J\cos\left(  \frac{2\pi k}{N}\right)  +\left(  N-4\right)  J_{z}.
\end{array}
\right.  \label{g6}%
\end{equation}

When the $XXZ$ chain couple with reservoir as \eqref{d23}, the evolution is no
longer unitary and this dynamic process can be described by the master
equation:%
\begin{equation}
\frac{d\rho}{dt}=-i\left[  H_{S},\rho\right]  +\gamma{\sum
\limits_{i=1}^{N}}\left(  2\sigma_{i}^{-}\rho\sigma_{i}^{+}-\sigma_{i}%
^{+}\sigma_{i}^{-}\rho-\rho\sigma_{i}^{+}\sigma_{i}^{-}\right)  \label{g7}%
\end{equation}
Through the previous models Eq. (\ref{d26},\ref{d34}) can be solved by $c$-number or other
method, this model \eqref{g7} is still unsolved. To solve the master equation,
we construct the corresponding Ket-Bra Entangled State%
\begin{equation}
|\eta_{S}\rangle={\sum\limits_{i=1}^{N}}{\sum\limits_{S_{i}=0,1}}|S_{1}%
,S_{2},\cdots,S_{N},\tilde{S}_{1},\tilde{S}_{2},\cdots,\tilde{S}_{N}%
\rangle\label{g8}%
\end{equation}
where $\tilde{S}_{i}=S_{i}=0,1$ represents the spin of $i$-th qubit. Similar
with \eqref{d14}, one can found%
\begin{equation}%
\begin{array}
[c]{l}%
\rho|\eta_{S}\rangle={\sum}\rho_{\varphi,\psi}|\varphi,\tilde{\psi}%
\rangle\equiv|\rho\rangle,\\
\sigma_{i}^{-}|\eta_{S}\rangle=\tilde{\sigma}_{i}^{+}|\eta_{S}\rangle
,\sigma_{i}^{+}|\eta_{S}\rangle=\tilde{\sigma}_{i}^{-}|\eta_{S}\rangle.
\end{array}
\label{g9}%
\end{equation}
Thus \eqref{g7} can be converted into Schr\"{o}dinger-like equation:%
\begin{equation}
\frac{d}{dt}|\rho\rangle=\mathscr{F}_{S}|\rho\rangle=\left(  H_{F}+G\right)
|\rho\rangle, \label{g10}%
\end{equation}
where
\begin{equation}%
\begin{array}
[c]{l}%
G=2\gamma{\sum\limits_{i}}\sigma_{i}^{-}\tilde{\sigma}_{i}^{-},\\
H_{F}=i\left(  \tilde{H}_{S}-H_{S}\right)  -\gamma{\sum\limits_{i}}\left(
\sigma_{i}^{+}\sigma_{i}^{-}+\tilde{\sigma}_{i}^{+}\tilde{\sigma}_{i}%
^{-}\right)  .
\end{array}
\label{g11}%
\end{equation}
The formal solution of \eqref{g10} is%
\begin{equation}
|\rho\left(  t\right)  \rangle=e^{\mathscr{F}_{S}t}|\rho\left(  0\right)
\rangle=e^{\left(  H_{F}+G\right)  t}|\rho\left(  0\right)  \rangle.
\label{g12}%
\end{equation}

Introduce the two-mode vector of $\vert k\rangle$%
\begin{equation}
\vert k,\tilde{k}^{\prime}\rangle=\vert k\rangle\langle k^{\prime}\vert
\vert\eta_{S}\rangle=\frac{1}{N}{\sum\limits_{m,n=1}^{N}}e^{\frac{i2\pi\left(
nk-mk^{\prime}\right)  }{N}}\sigma_{n}^{+}\tilde{\sigma}_{m}^{+}\vert
0,\tilde{0}\rangle. \label{g13}%
\end{equation}
Easy to found%
\begin{equation}%
\begin{array}
[c]{ll}%
G\vert0,\tilde{0}\rangle=0, & H_{F}\vert0,\tilde{0}\rangle=0;\\
G\vert0,\tilde{k}\rangle=0, & H_{F}\vert0,\tilde{k}\rangle=h_{0,k}%
\vert0,\tilde{k}\rangle;\\
G\vert k,\tilde{0}\rangle=0, & H_{F}\vert k,\tilde{0}\rangle=h_{0,k}^{\ast
}\vert k,\tilde{0}\rangle;\\
G\vert k,\tilde{k}^{\prime}\rangle=g_{k,k^{\prime}}\vert0,\tilde{0}\rangle, &
H_{F}\vert k,\tilde{k}^{\prime}\rangle=h_{k,k^{\prime}}\vert k,\tilde
{k}^{\prime}\rangle;
\end{array}
\label{g14}%
\end{equation}
where
\begin{equation}%
\begin{array}
[c]{l}%
g_{k,k^{\prime}}=2\gamma\delta_{k,k^{\prime}},\\
h_{0,k}=i\left(  E_{k}-E_{0}\right)  -\gamma,\\
h_{k,k^{\prime}}=i\left(  E_{k^{\prime}}-E_{k}\right)  -2\gamma.
\end{array}
\label{g15}%
\end{equation}
With \eqref{g14}, for $\vert0,\tilde{k}\rangle$ and $\vert k,\tilde{0}\rangle$
we obtain%
\begin{equation}%
\begin{array}
[c]{l}%
\vert\rho_{0,k}\left(  t\right)  \rangle=e^{\left(  H_{F}+G\right)  t}%
\vert0,\tilde{k}\rangle=e^{h_{0,k}t}\vert0,\tilde{k}\rangle\\
\vert\rho_{k,0}\left(  t\right)  \rangle=e^{\left(  H_{F}+G\right)  t}\vert
k,\tilde{0}\rangle=e^{h_{0,k}^{\ast}t}\vert k,\tilde{0}\rangle
\end{array}
\label{g16}%
\end{equation}
while for $\vert k,\tilde{k}^{\prime}\rangle$
\begin{align}
\vert\rho_{k,k^{\prime}}\left(  t\right)  \rangle &  =e^{\left(
H_{F}+G\right)  t}\vert k,\tilde{k}^{\prime}\rangle=\sum_{n=0}^{\infty}%
\frac{t^{n}}{n!}\left(  H_{F}^{n}+GH_{F}^{n-1}\right)  \vert k,\tilde
{k}^{\prime}\rangle,\nonumber\\
&  =e^{h_{k,k^{\prime}}t}\vert k,\tilde{k}^{\prime}\rangle+\frac
{g_{k,k^{\prime}}}{h_{k,k^{\prime}}}\left(  e^{h_{k,k^{\prime}}t}-1\right)
\vert0,\tilde{0}\rangle. \label{g17}%
\end{align}

Consider the initial state is $\rho\left(  0\right)  =\vert k\rangle\langle
k\vert$ i.e. $\vert\rho\left(  0\right)  \rangle=\vert k,\tilde{k}\rangle$
according to \eqref{g15} and \eqref{g17},
\begin{equation}
\vert\rho\left(  t\right)  \rangle=e^{-2\gamma t}\vert k,\tilde{k}%
\rangle+\left(  1-e^{-2\gamma t}\right)  \vert0,\tilde{0}\rangle. \label{g18}%
\end{equation}
Moreover, fix the initial state $\rho\left(  0\right)  =|\varphi\rangle
\langle\varphi|,|\varphi\rangle=a|0^{\otimes N}\rangle+b\sigma_{1}%
^{+}|0^{\otimes N}\rangle,$ where $a$ is real and $a^{2}+|b|^{2}=1$. Accodring
to \eqref{g4} and \eqref{g9}, the solution can be expressed as%
\begin{align}
|\rho\left(  0\right)  \rangle &  =\left(  |b|^{2}\sigma_{1}^{+}\tilde{\sigma
}_{1}^{+}+ab\sigma_{1}^{+}+ab^{\ast}\tilde{\sigma}_{1}^{+}+a^{2}\right)
|0,\tilde{0}\rangle,\nonumber\\
&  =\frac{|b|^{2}}{N}{\sum\limits_{k,k^{\prime}=1}^{N}}e^{\frac{i2\pi\left(
k^{\prime}-k\right)  }{N}}|k,\tilde{k}^{\prime}\rangle+a^{2}|0,\tilde
{0}\rangle\nonumber\\
&  +\frac{ab}{\sqrt{N}}{\sum\limits_{k=1}^{N}}e^{\frac{-i2\pi k}{N}}%
|k,\tilde{0}\rangle+\frac{ab^{\ast}}{\sqrt{N}}{\sum\limits_{k=1}^{N}}%
e^{\frac{i2\pi k}{N}}|0,\tilde{k}\rangle\label{g19}%
\end{align}
As what we have done in \eqref{g16} and \eqref{g17}, the solution is given by
\begin{align}
|\rho\left(  t\right)  \rangle=  &  \left(  1-|b|^{2}e^{-2\gamma t}\right)
|0,\tilde{0}\rangle+\frac{ab}{N}{\sum\limits_{n,k=1}^{N}}e^{\frac{i2\pi\left(
n-1\right)  k}{N}+h_{0,k}^{\ast}t}\sigma_{n}^{+}|0^{\otimes N},\tilde
{0}^{\otimes N}\rangle\nonumber\\
&  +\frac{|b|^{2}}{N^{2}}{\sum\limits_{m,n,k,k^{\prime}=1}^{N}}e^{\frac
{i2\pi\left[  \left(  n-1\right)  k-\left(  m-1\right)  k^{\prime}\right]
}{N}+h_{k,k^{\prime}}t}\sigma_{n}^{+}\tilde{\sigma}_{m}^{+}|0^{\otimes
N},\tilde{0}^{\otimes N}\rangle\nonumber\\
&  +\frac{ab^{\ast}}{N}{\sum\limits_{n,k=1}^{N}}e^{\frac{-i2\pi\left(
n-1\right)  k}{N}+h_{0,k}t}\tilde{\sigma}_{n}^{+}|0^{\otimes N},\tilde
{0}^{\otimes N}\rangle. \label{g20}
\end{align}
Trace to other qubit except $j$-qubit, the density of $j$-qubit is obtained
\begin{equation}%
\begin{array}
[c]{l}
\rho_{01}^{j}\left(  t\right)  =\frac{ab^{\ast}}{N}{\sum\limits_{k=1}^{N}%
}e^{\frac{-i2\pi\left(  j-1\right)  k}{N}+h_{0,k}t}\\
\rho_{11}^{j}\left(  t\right)  =\frac{\vert b\vert^{2}}{N^{2}}{\sum
\limits_{k,k^{\prime}=1}^{N}}e^{\frac{i2\pi\left(  j-1\right)  \left(
k-k^{\prime}\right)  }{N}+h_{k,k^{\prime}}t}%
\end{array}
\label{g21}%
\end{equation}
other elements can be given by $\rho_{00}^{j}\left(  t\right)  =1-\rho
_{11}^{j}\left(  t\right)  $ and $\rho_{10}^{j}\left(  t\right)  =\rho
_{01}^{j\ast}\left(  t\right)  $. To analysis the decoherence evolution of the
Heisenberg chain, we plot $\rho_{11}^{j}\left(  t\right)  $ and $\vert
\rho_{01}^{j}\left(  t\right)  \vert$ as functions of the dimensionless time
$\gamma t$

\begin{figure}[h]
\setcounter{subfigure}{0} \centering
\subfigure[\ $N=3$]{
\includegraphics[width=0.23\textwidth]{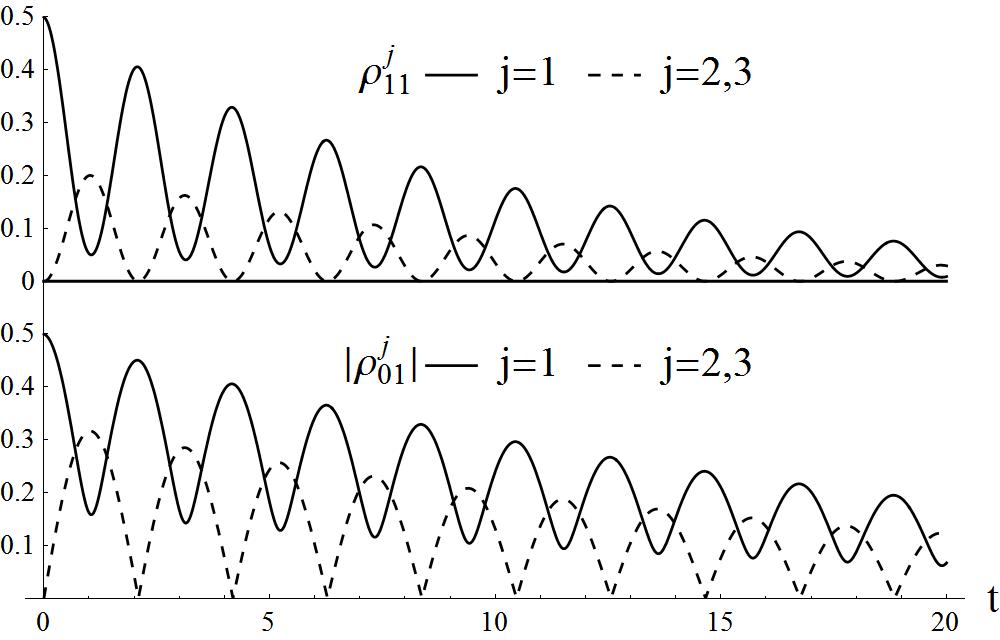}} \hspace{0.02in}
\subfigure[$\  N=4$]{
\includegraphics[width=0.23\textwidth]{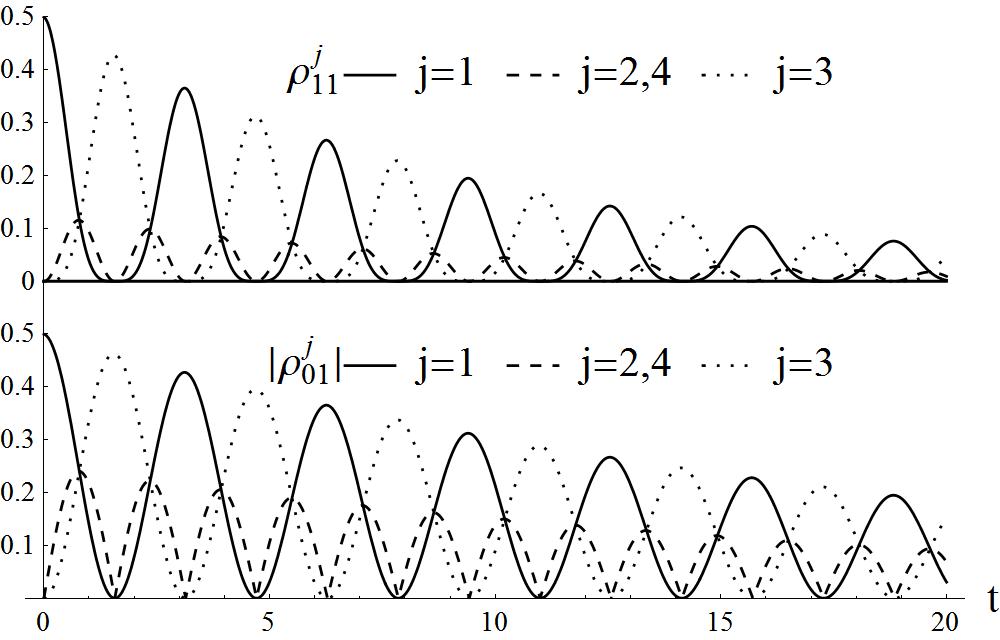}}\newline%
\subfigure[$\  N=5$]{
\includegraphics[width=0.23\textwidth]{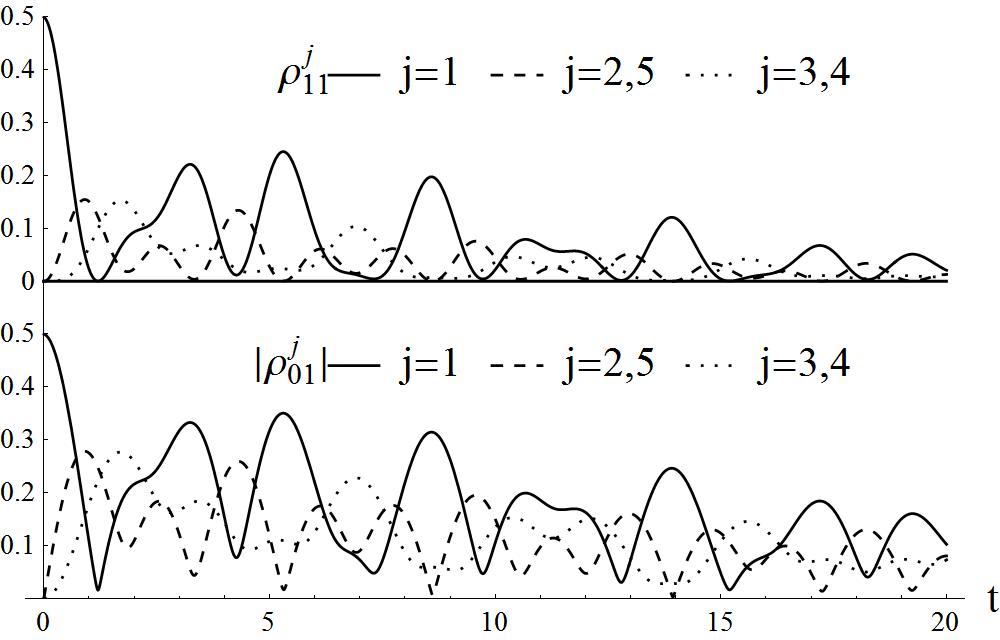}} \hspace{0.02in}
\subfigure[$\  N=6$]{
\includegraphics[width=0.23\textwidth]{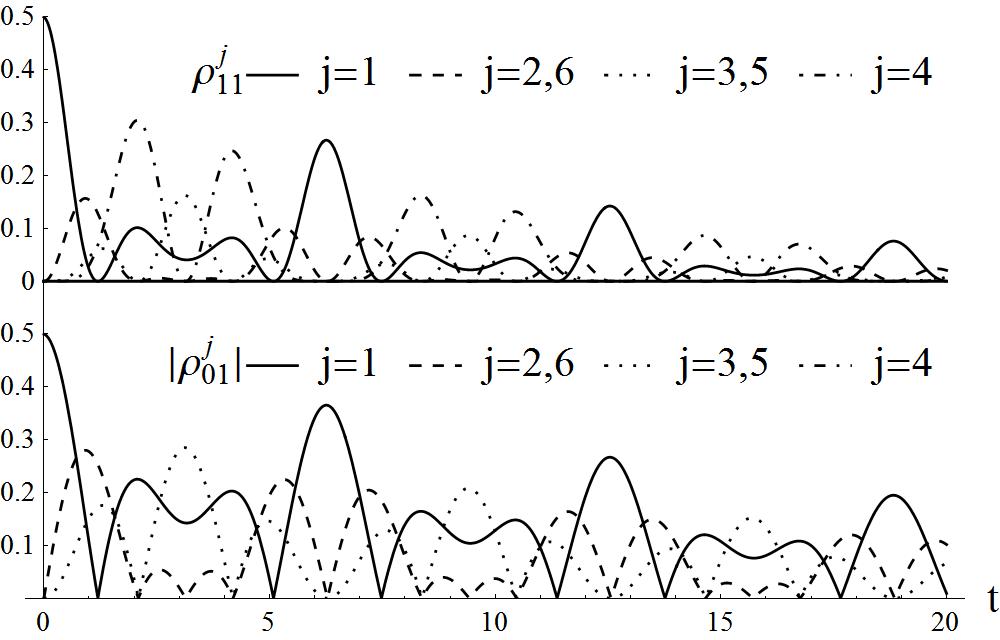}}
\caption{$\rho_{11}^{j}\left(  t\right)$ and $\vert\rho_{01}^{j}\left(  t\right)  \vert$ when $J=2,\gamma=1/220,a=\left\vert b\right\vert=1/\sqrt{2}$.}
\end{figure}

In Fig.5, the evolution of $\rho_{11}^{j}\left(  t\right)  $ and
$\vert\rho_{01}^{j}\left(  t\right)  \vert$ is no longer monotone decreasing
but fluctuate with different frequency, the frequency increases with the
growth of $J/\gamma$. Fig.5 show that there are some cross points for $N=3,4$,
while $N=5,6$ is not; Moreover, the cross points only occurs when $N\leq
4$, and disappear for $N\geq5$.

For the case $N=4$, the abscissa of cross points can be given by%
\begin{equation}
\rho_{11}^{1}\left(  t\right)  =\rho_{11}^{2}\left(  t\right)  =\rho_{11}%
^{3}\left(  t\right)  =\rho_{11}^{4}\left(  t\right)  .\label{g22}%
\end{equation}
The solution is given by
\begin{equation}
Jt_{n}^{\pm}=2n\pi\pm\frac{\pi}{2}\label{g23}%
\end{equation}
Substituting the value of $t_{n}^{\pm}$ into \eqref{g20}, the density $\rho$ at
cross points is%
\begin{align}
&  \rho\left(  t_{n}^{\pm}\right)  =|\psi\left(  t_{n}^{\pm}\right)
\rangle\langle\psi\left(  t_{n}^{\pm}\right)  |+|b|^{2}\left(  1-e^{-2\gamma
t_{n}^{\pm}}\right)  |0^{\otimes4}\rangle\langle0^{\otimes4}|,\nonumber\\
&  |\psi\left(  t_{n}^{\pm}\right)  \rangle\equiv be^{-\gamma t_{n}}%
|\varphi^{\pm}\rangle+ae^{-4iJ_{z}t_{n}^{\pm}}|0^{\otimes4}\rangle,\label{g24}%
\end{align}
where $|\varphi^{\pm}\rangle$ is the $W$-states for $4$ qubit
\begin{equation}
|\varphi^{\pm}\rangle=\frac{1}{2}\left(  |1000\rangle\pm i|0100\rangle
-|0010\rangle\pm i|0001\rangle\right)  .\label{g25}%
\end{equation}
That means $\rho\left(  t_{n}^{\pm}\right)  $ consist of two
parts, one is all spin down states $|0^{\otimes4}\rangle$, the other is
$|\psi\left(  t_{n}^{\pm}\right)  \rangle$ a superposition state of $W$-states
$|\varphi^{\pm}\rangle$ and $|0^{\otimes4}\rangle$. $\rho\left(  t_{n}^{\pm
}\right)  $ reduced to a pure $W$-states $|\varphi^{\pm}\rangle\langle
\varphi^{\pm}|$ for $b=1,\gamma=0$. \eqref{g23} exhibit that the frequency of 
cross points $f_{cp}=J/\pi$ is independent of $\gamma$, while \eqref{g23} 
show that $\gamma$ only influence the exponential damping of both functions.
All these imply that the occur of cross point and $W$-state should be attribute
to the interact of Heisenberg model, $\gamma$ only effects on dissipation process.

In order that a certain state occurs periodically in the system, the necessary
and unsufficient condition is that the ratio of any two frequencies is a
rational number. \eqref{g6} show that the condition is satisfied for $2,3,4$ and
$6$ qubits. For $6$-qubits the $\rho_{11}^{j}\left(  t\right)  $ evolve
 decaying periodically with time, however no cross point exist namely $6$-qubit
$W$-states can't be created by this model. All the solution \eqref{g20},
$W$-states\eqref{g25}, and entanglement evolution still need further
investigation. However, in present paper our primary purpose is to introduce
our new Ket-Bra Entangled State method, so we shall research those problems in
the near future.
\section{Conclusion}

In this paper we present a new method (Sec. II) that map a master equation
into a Schr\"{o}dinger-like equation, so most procedure of Schr\"{o}dinger
equation can be used to solve the master equation. All master equation of
finite dimension system can be resolved by this KBES method in theory. To
solve master equation of $N$-level system, the calculation of $N^{2}$-order
matrix's exponent is necessary (see \eqref{d38}), whereas this tedious matrix
operation can be finished by computer. For other special cases, the way of
stationary Sch\"{o}rdinger equation method may simplify the calculation effectively.

Through this method, we solve the model of a damped qubit in time-dependent
external field and a qutrit coupled to reservoir,
then we resolve $N$-qubit Heisenberg chain each coupled with reservoir 
at zero temperature, and preliminarily analyze the dissipation
dynamics and decoherence dynamics find that $W$-states plays an important role
in this process (see Sec.III.c). All these cases show that KBES method is a
generalization and systematization method for solving master equation, which
can greatly oversimplify the resolution of master equation.

\end{document}